\documentstyle[preprint,revtex]{aps}
\begin{document}
\draft
\preprint{RU96--2--B; ETH-Lausanne}
\preprint{February 1996}
\begin{title}
Integrable $SU(m|n)$ supersymmetric electronic models\\ 
of strong correlations 
\end{title}
\author{James T. Liu}
\begin{instit}
Department of Physics\\
The Rockefeller University\\
1230 York Avenue\\
New York, NY 10021-6399, USA\\
\end{instit}
\author{D. F. Wang}
\begin{instit} 
Institut de Physique Th\'eorique\\
Ecole Polytechnique F\'ed\'erale de Lausanne\\
PHB-Ecublens, CH-1015 Lausanne, Switzerland
\end{instit}
\begin{abstract}
We generalize the $SU(2|2)$ symmetric extended Hubbard model of $1/r^2$
interaction to the $SU(m|n)$ supersymmetric case.  Integrable models may be
defined on both uniform and non-uniform one dimensional lattices.  We study
both cases in detail and present the ground state wavefunctions and energy
spectra of these models.
\end{abstract}
\narrowtext

\section{Introduction}
Much progress has been made in the study of low dimensional systems of
strongly correlated electrons.  A notable example is the one dimensional
Hubbard model, which was solved by Lieb and Wu\cite{wu} using the
Bethe-ansatz.  This model exhibits a hidden $SO(4)$ symmetry\cite{yang1,yang2}
which is the basis of its properties.  The suggestion of Anderson that 
the two dimensional Hubbard model contains the essential physics of high
temperature superconductivity and that its normal state may share the
Luttinger-liquid-like feature of one dimensional interacting electron
systems\cite{anderson,su} has induced renewed activity in the study
of such strongly correlated one dimensional models.

Recently, an extended Hubbard model with nearest neighbor interactions
was introduced\cite{korepin1,korepin2}, and is solvable by Bethe-ansatz
in one dimension \cite{korepin3}.  The spin $SU(2)$ along with the
$\eta$-pairing $SU(2)$ of
this model are combined and extended to form a $SU(2|2)$ supersymmetry.
We have introduced the Calogero-Sutherland\cite{calogero,sutherland} version
of the extended Hubbard model, defined both on a uniform lattice and on a
non-uniform lattice in one dimension.  The long range versions of the
extended Hubbard model are exactly solvable, and the ground states,
excitations and integrabilities have been studied in considerable
detail\cite{wang4,liu}. 

In this paper, we generalize the previous results by extending the
supersymmetry to $SU(m|n)$.  Thus we introduce the $SU(m|n)$ strongly
correlated electron system of $1/r^2$ long range interaction.  This
general model encompasses the $SU(0|2)$ Haldane-Shastry spin
chain\cite{haldane,shastry} and the $SU(1|2)$ supersymmetric $t$-$J$ model 
\cite{kura} as well as the $SU(2|2)$ extended Hubbard model.

In the next section we present the general features of the $SU(m|n)$
supersymmetric model.  Following this, we study the system, first defined
on a non-uniform lattice given by the roots of the Hermite polynomials, and
then defined on a uniform lattice.  In both cases, the $SU(m|n)$ models
are completely solvable.  We explicitly construct the ground state
wavefunctions and provide the ground state energies.  We also examine the
excitation spectra of these systems, making use of their integrability
properties.

\section{The $SU(m|n)$ supersymmetric electronic model}

Consider a system defined on a lattice with a total of $L$ sites.  At each
site there is exactly one particle, which may either be a fermion with $n$
flavors (internal degrees of freedom) or a boson with $m$ flavors.  Since
there are a total of $m+n$ states at each site, $n$ fermionic and $m$ bosonic,
this is the basis of the $SU(m|n)$ supersymmetry.

In order to describe these degrees of freedom, we use a representation in
terms of fermion and boson operators, $f$ and $b$, satisfying the usual
(anti-)commutation relations
\begin{eqnarray}
\{f_{i\sigma}^{\vphantom{\dagger}},f_{j\sigma'}^\dagger\}
&=&\delta_{ij}\delta_{\sigma\sigma'}
\qquad \{f_{i\sigma},f_{j\sigma'}\}=0\nonumber\\
{}[b_{i\alpha}^{\vphantom{\dagger}}, b_{j\alpha'}^\dagger]
&=&\delta_{ij} \delta_{\alpha\alpha'}
\qquad [b_{i\alpha},b_{j\alpha'}]=0\ .
\end{eqnarray}
The indices $i$ and $j$ label the site, $\sigma=1,\ldots,n$ labels the
fermion flavor and $\alpha=1,\ldots,m$ labels the boson flavor.  These
operators may be represented in a supersymmetric manner by
$c_{i\nu}=\{f_{i\sigma}\},\{b_{i\alpha}\}$ where $\nu=1,\ldots,m+n$ labels
the (either fermionic or bosonic) species.  The $c$'s then satisfy the
(anti-)commutation relation
\begin{equation}
[c_{i\nu}^{\vphantom{\dagger}},c_{j\nu'}^\dagger]_\pm
=\delta_{ij}\delta_{\nu\nu'}\ ,
\end{equation}
where $[\ldots]_\pm$ indicates either a commutator or an anti-commutator
as appropriate.  In general, this representation allows for multiple
occupancy at each site.  In order to project onto single occupancy we use
the Gutzwiller operator, $P_G$, defined by
\begin{eqnarray}
P_G&=&\prod_{i=1}^L \delta_{1,\sum_\nu c_{i\nu}^\dagger
    c_{i\nu}^{\vphantom{\dagger}}}\nonumber\\
&=&\prod_{i=1}^L \delta_{1,
  \sum_\sigma f_{i\sigma}^\dagger f_{i\sigma}^{\vphantom{\dagger}} +
  \sum_\alpha b_{i\alpha}^\dagger b_{i\alpha}^{\vphantom{\dagger}}}\ .
\end{eqnarray}

The Hamiltonian for the $SU(m|n)$ supersymmetric model takes the form
\begin{equation}
H=-{1\over 2} P_G \sum_{1\le i\ne j\le L} J_{ij} \Pi_{ij}P_G\ ,
\label{eq:hamil}
\end{equation}
where the coupling parameter $J_{ij}=1/d(i,j)^2$ takes an inverse square
form.  The distance function, $d(i,j)$, depends on the lattice and is what
distinguishes between the models on uniform and non-uniform lattices.  In
this section, we concentrate on the features of $H$ that are independent of
the specifics of $d(i,j)$.  The key features of this model, especially its
integrability, are derived from the graded permutation operator, $\Pi_{ij}$,
which exchanges particle species at sites $i$ and $j$.  In the above
representation, the permutation operator takes the form
\begin{eqnarray}
\Pi_{ij}&=&\sum_{\nu=1}^{m+n}\sum_{\nu'=1}^{m+n}
c_{i\nu}^\dagger c_{j\nu'}^\dagger
   c_{i\nu'}^{\vphantom{\dagger}}c_{j\nu}^{\vphantom{\dagger}}\nonumber\\
&=&-\sum_{\sigma=1}^n\sum_{\sigma'=1}^n
f_{i\sigma}^\dagger f_{i\sigma'}^{\vphantom{\dagger}}
   f_{j\sigma'}^\dagger f_{j\sigma}^{\vphantom{\dagger}}
+ \sum_{\alpha=1}^m \sum_{\alpha'=1}^m
b_{i\alpha}^\dagger b_{i\alpha'}^{\vphantom{\dagger}}
   b_{j\alpha'}^\dagger b_{j\alpha}^{\vphantom{\dagger}}\nonumber\\
&&+ [\sum_{\sigma=1}^n \sum_{\alpha=1}^m
f_{i\sigma}^\dagger b_{i\alpha}^{\vphantom{\dagger}}
    b_{j\alpha}^\dagger f_{j\sigma}^{\vphantom{\dagger}} + h.c.]\ .
\label{eq:gperm}
\end{eqnarray}

Since the Hamiltonian only permutes particles, the number of bosons and
fermions of each flavor are all conserved quantities.  It is also easy to
verify that $H$ commutes with both the $SU(n)$ fermion and $SU(m)$ boson
flavor symmetries generated by
\begin{equation}
S^a=\sum_{i=1}^L S_i^a,\qquad
S_i^a=\sum_{\sigma,\sigma'}f_{i\sigma}^\dagger T^a_{\sigma\sigma'}(n)
f_{i\sigma'}^{\vphantom{\dagger}}\ ,
\label{eq:nfermion}
\end{equation}
and
\begin{equation}
\eta^a=\sum_{i=1}^L \eta_i^a,\qquad
\eta_i^a=\sum_{\alpha,\alpha'}b_{i\alpha}^\dagger T^a_{\alpha\alpha'}(m)
b_{i\alpha'}^{\vphantom{\dagger}}\ ,
\label{eq:mboson}
\end{equation}
respectively (here $T(n)$ and $T(m)$ are hermitian $SU(n)$ and $SU(m)$
generators in the fundamental representations).  In fact, ignoring the
grading, since
$\Pi_{ij}$ acts symmetrically on all $m+n$ species [as is evident from
the first line of (\ref{eq:gperm})], we would expect a $SU(m+n)$ symmetry
of this model.  Taking statistics into account, this is modified into a
$SU(m|n)$ supersymmetry of which the bosonic subgroup is given by $SU(m)\times
SU(n)\times U(1)$ (see {\it e.g.}\ Ref.~\cite{hurni}).  The $U(1)$ symmetry
is related to the boson and fermion number currents, and is given by
\begin{equation}
J=\sum_{i=1}^L J_i,\qquad
J_i={1\over n}\sum_{\sigma=1}^n
   f_{i\sigma}^\dagger f_{i\sigma}^{\vphantom{\dagger}}
+{1\over m}\sum_{\alpha=1}^m
   b_{i\alpha}^\dagger b_{i\alpha}^{\vphantom{\dagger}}\ .
\label{eq:u1}
\end{equation}
Although it appears there is another $U(1)$ symmetry given by the remaining
orthogonal combination of boson and fermion currents, this is actually not
the case since we restrict ourselves to single occupancy, and thus total
particle number is fixed.  This means that the second $U(1)$ symmetry is
trivial (except in the special case when $m=n$).

Complementing the above bosonic generators of $SU(m|n)$ are the fermionic
generators
\begin{eqnarray}
Q_{\sigma\alpha}&=&\sum_{i=1}^L Q_{i\sigma\alpha},\qquad
Q_{i\sigma\alpha}=f_{i\sigma}^\dagger b_{i\alpha}^{\vphantom{\dagger}}
\ ,\nonumber\\
Q_{\alpha\sigma}^\dagger&=&\sum_{i=1}^L Q_{i\alpha\sigma}^\dagger,\qquad
Q_{i\alpha\sigma}^\dagger=b_{i\alpha}^\dagger
f_{i\sigma}^{\vphantom{\dagger}}\ ,
\end{eqnarray}
transforming as the $(\overline{m},n)$ and $(m,\overline{n})$ of
$SU(m)\times SU(n)$ respectively.  The complete $SU(m|n)$ superalgebra is
given by these $2mn$ fermionic generators as well as the $m^2+n^2-1$
bosonic generators of Eqns.~(\ref{eq:nfermion}), (\ref{eq:mboson}) and
(\ref{eq:u1}).

In general, since the Hamiltonian $H$ is $SU(m|n)$ invariant, all states in
the system fall into complete representations of the $SU(m|n)$ supersymmetry
algebra.  These {\it supermultiplets} have a very rich structure, and span
states with different numbers of fermions and bosons.  Representations of
$SU(m|n)$ may be divided into two categories, ``typical'' and ``atypical''
\cite{hurni}.  The typical representations group together states in
sectors with boson and fermion numbers given by
$(Q,M)=(b,f),(b+1,f-1),(b+2,f-2), \ldots, (b+mn,f-mn)$, while the
atypical ones terminate early and hence span fewer than the $mn+1$
sectors of typical ones.  In both cases, a full $SU(m|n)$ representation
may be decomposed in terms of a direct sum (over sectors of decreasing
fermion number) of representations of the bosonic subgroup
$SU(m)\times SU(n)\times U(1)\subset SU(m|n)$.  For an irreducible
representation, we refer to the state with greatest fermion
number ({\it i.e.}~the $(b,f)$ sector) as the highest member of a
supermultiplet and the state with the fewest fermions (the $(b+mn,f-mn)$
sector for typical representations) as the lowest member.

While the important properties of the system may be understood through this
supermultiplet structure, we generally wish to have greater control over
the individual species populating the lattice.  In particular, we note that
since $SU(m|n)$ has a maximal Cartan subalgebra given by $U(1)^{m+n-1}$, 
there are a set of $m+n-1$ conserved currents which are equivalent to the
fermion and boson number currents of the $m+n$ species (along with the
single occupancy constraint).  As a result, the number of bosons of each
flavor and the number of the fermions of each flavor are all conserved
quantities, and hence we may work in a subspace where all these quantities
are fixed.  This is easily accomplished by adding to $H$ a set of chemical
potentials for the different species:
\begin{equation}
{\cal H}=H+\sum_{\nu=1}^{m+n} \mu^{(\nu)}\sum_{i=1}^Lc_{i\nu}^\dagger
c_{i\nu}^{\vphantom{\dagger}}\ ,
\end{equation}
(one of the $\mu$'s is redundant).  
Each subspace of fixed occupancies may be labeled by
the $m+n$ numbers $\{Q_\alpha,M_\sigma\}$ given by
\begin{eqnarray}
&&M_{\sigma}=\sum_{i=1}^L f_{i\sigma}^\dagger f_{i\sigma}^{\vphantom{\dagger}},
\qquad\sigma=1, 2, \cdots,n\nonumber\\
&&Q_{\alpha}=\sum_{i=1}^L b_{i\alpha}^\dagger b_{i\alpha}^{\vphantom{\dagger}},
\qquad\alpha=1,2, \cdots,m\ .
\end{eqnarray}
Since each site is singly occupied, these quantities satisfy the
relation that $\sum_\sigma M_\sigma+\sum_\alpha Q_\alpha=L$.  Working in
such a subspace, we have broken the $SU(m|n)$ symmetry down to its maximal
Cartan subalgebra.  Alternatively, we often consider the case when $SU(m|n)$
is not broken completely down to $U(1)^{m+n-1}$, but instead to its bosonic
subgroup $SU(m)\times SU(n)\times U(1)$.  This is equivalent to working in
a subspace of fixed $(Q,M)$ where
$Q=\sum_\alpha Q_\alpha$ is the total number of bosons and
$M=\sum_\sigma M_\sigma$ is the total number of fermions, and corresponds
to using the $U(1)$ current $J$ to break supersymmetry.

We now define the wavefunctions in this Hilbert subspace of fixed
$(\{Q_\alpha\},\{M_\sigma\})$ by writing the state vectors of the system
in the following form:
\begin{equation}
|\phi\rangle=\sum_{\{x\sigma\}, \{y\alpha\}} \phi(x\sigma; y\alpha) 
\prod_{i=1}^M f_{x_i\sigma_i}^\dagger \prod_{k=1}^Q b_{y_k\alpha_k}^\dagger
|0\rangle\ .
\label{eq:amplitude}
\end{equation}
The amplitude $\phi(x\sigma,y\alpha)$ is antisymmetric in the fermion
positions and spins, $\{x\sigma\}=(x_1\sigma_1,\cdots,x_M\sigma_M)$, while
symmetric in the boson positions and spins, 
$\{y\alpha\}=(y_1\alpha_1, \cdots, y_Q\alpha_Q)$.  Because of the
supersymmetry, it is convenient to combine the boson and fermion positions
by defining $(q_1, q_2, \cdots, q_L)=(x_1, \cdots, x_M|y_1,\cdots,y_Q)$
and $\phi(x\sigma,y\alpha)=\phi(\{q\},\{\sigma\},\{\alpha\})$.  Due to
single occupancy, the set $\{q\}$ spans the lattice.

With the state vector written in the above manner, the fermionic and
bosonic nature of
the particles is encoded in the symmetry properties of the wavefunction.
Thus the graded permutation operator, (\ref{eq:gperm}), takes a particularly
simple form independent of the particle statistics.  The resulting
Hamiltonian, acting on wavefunctions $\phi(\{q\},\{\sigma\},\{\alpha\})$, is
\begin{equation}
H=-{1\over2}\sum_{1\le i\ne j\le L} {M_{ij}\over d(q_i,q_j)^2}\ ,
\label{eq:phihamil}
\end{equation}
where $M_{ij}$ interchanges particles $i$ and $j$,
$M_{ij} \phi(\{q\},\{\sigma\},\{\alpha\})=\phi(\{q'\},\{\sigma\},\{\alpha\})$, 
with $\{q'\}=(q_1',q_2',\cdots,q_L')=(q_1,\cdots,q_j,\cdots,q_i,\cdots,q_L)$. 
The resulting eigenenergy equation then takes the form
\begin{equation}
-{1\over 2} \sum_{1\le i\ne j\le L} d(q_i,q_j)^{-2} M_{ij} 
\phi (\{q\},\{\sigma\},\{\alpha\}) 
=E \phi(\{q\},\{\sigma\},\{\alpha\})\ .
\label{eq:eigen}
\end{equation}

We immediately see that the lowest energy state in the full Hilbert space
corresponds to constant $\phi$.  Since this is only compatible with bosonic
symmetry, we conclude that the $SU(m|n)$ ground state is in the $M=0$ sector
and is described by the wavefunction
\begin{equation}
\phi_G (y\alpha) = 1\ .
\label{eq:fullgs}
\end{equation} 
Because this is a good wavefunction independent of boson flavors $\alpha$,
the ground state is in general degenerate, and is given by the $L$-fold
symmetric combination of the fundamental representation of $SU(m)$.  In
terms of the Young tableau, this is the $([L^1],[\cdot])$ of $SU(m)\times
SU(n)$.  Since the wavefunction is constant, the ground state energy is
given by
\begin{equation}
E_G=-{1\over2}\sum_{1\le i\ne j\le L}d(i,j)^{-2}\ ,
\end{equation}
and may easily be summed.  All excitations, whether fermionic or bosonic,
are built on top of this.

To be more precise about the nature of the ground state, the purely
bosonic $(Q,M)=(L,0)$ state is just the lowest member of a $SU(m|n)$
supermultiplet,
and therefore its degeneracy in the full Hilbert space is larger than
indicated by the $M=0$ sector alone.  We find that the ground state
representation is atypical, and spans the sectors $(L-n,n),(L-n+1,n-1),
\ldots,(L,0)$ with corresponding representations $([(L-n)^1],[1^n])$,
$([(L-n+1)^1],[1^{n-1}]),\ldots,([L^1],[\cdot])$.  Alternatively, this
result follows by noting that fermionic wavefunctions which are symmetric in
the coordinates $\{x\}$ may be constructed by taking antisymmetric
combinations of their flavors $\sigma$.

The highest energy state of the $SU(m|n)$ model is given by a completely
antisymmetric wavefunction, corresponding to the fully fermionic sector,
$M=L$.  In the absence of periodic boundary conditions, the wavefunction is
\begin{equation}
\phi_M (x\sigma) = \prod_{i<j} (q_i-q_j)\ ,
\end{equation}
with energy $E_M=-E_G$.  Since $\phi_M$ is independent of the fermion
flavor $\sigma$, the highest energy state corresponds to the
$([\cdot],[L^1])$ representation (this time the highest member of a
supermultiplet).  As a result, all states in the spectrum
lie in the energy range $[E_G,-E_G]$.  Furthermore, any eigenstate of $H$,
when multiplied by $\phi_M$, remains a good eigenstate, but with opposite
eigenenergy.  Since this also interchanges bosonic and fermionic boundary
conditions, it corresponds to the interchange of the $SU(m|n)$ and $SU(n|m)$
theories.  Therefore this indicates that $H$ for the $SU(m|n)$ model
corresponds to $-H$ for the $SU(n|m)$ model \cite{haldanetalk}.  We will
subsequently make use of this symmetry to obtain the upper bound of the
energy levels of the $SU(m|n)$ permutation model defined on the Hermite
lattice.

In general, we wish to work, not in the full Hilbert space, but rather in a
given subspace of fixed $(Q,M)$.  In this case, wavefunctions have mixed
symmetry properties.  We will consider the specifics of these wavefunctions
in the next two sections.  Here we mention that, since the ground state in
this sector contains a symmetric combination of $Q$ bosons, it is again
degenerate \cite{sutherland2}, and corresponds to the $Q$-fold symmetric
combination of the $m$ of $SU(m)$.  As a general rule, in order to lower the
energy, the wavefunction should be as symmetric as possible.  This may be
accomplished by distributing the $M$ fermions as equally as possible into
the $n$ flavors, resulting in the ground state representation
$([Q^1],[1^r])$ where $r = M\,{\rm mod}\, n$.  This will be made clear
subsequently for both the non-uniform and the uniform lattice.

\section{$SU(m|n)$ electronic model on a non-uniform lattice}

As we have seen above, many properties of the $SU(m|n)$ model are quite
general, and result from the permutation nature of the system.
In order to proceed further with a detailed study of the excitation
spectrum, we now fix the lattice, as specified by the distance function
$d(i,j)$.  In this section we discuss the $SU(m|n)$ model defined on
a non-uniform lattice in one dimension where the sites of the 
chain are given by the roots of the Hermite polynomial $H_L(x)$. 
It is well known that this Hermite polynomial has $L$ roots,
$r_1, r_2, \ldots, r_L$, which are all real and distinct.  Thus the distance
function for this non-uniform lattice is defined by
\begin{equation}
d(i,j) = |r_i-r_j|\ .
\end{equation}
This one dimensional chain is well defined and yields an exactly integrable
system.  The $SU(m|n)$ model on a non-uniform lattice encompasses
the $SU(0|n)$ spin chain with inverse square exchange \cite{poly} and the
$SU(1|n)$ supersymmetric $t$-$J$ model\cite{wang1,wang2} as well as the
$SU(2|2)$ extended Hubbard model\cite{wang4}.

Since the Hamiltonian, (\ref{eq:phihamil}), is written solely in terms of
permutations on the particle positions and is independent of statistics,
many results for the $SU(0|2)$ spin chain on a non-uniform lattice
\cite{poly} are applicable to the general $SU(m|n)$ system as well.  In
particular, the Hamiltonian commutes with a set of conserved quantities
$\{I_A\}$, namely
\begin{equation}
[H,I_A]=0,\qquad[I_A,I_B]=0,\qquad A,B=0,1,2, \ldots, \infty\ ,
\end{equation}
where
\begin{eqnarray}
&&I_A=\sum_{i=1}^L (a_i^\dagger a_i^{\vphantom{\dagger}})^A
\nonumber\\
&&a_k^\dagger = i\sum_{j(\ne k)=1}^L (q_k-q_j)^{-1}M_{kj} + iq_k,
\qquad a_k^{\vphantom{\dagger}}=(a_k^\dagger)_{\vphantom{k}}^\dagger\ .
\end{eqnarray}
These relations yield the integrability conditions of the $SU(m|n)$ model.
The quantities $I_A$ as written act only on the coordinates of wavefunctions
$\phi(\{q\},\{\sigma\}, \{\alpha\})$ and thus commute with the $SU(m|n)$
algebra.  One can use the permutation symmetry property of the wavefunction
$\phi$ defined by Eqn.~(\ref{eq:amplitude}) in order to write explicit
representations for $I_A$ in terms of the $b$ and $f$ operators.

As shown in the previous section, the ground state of this model in the
complete Hilbert space has its lowest component in the $(Q,M)=(L,0)$ subspace,
and has an energy
\begin{equation}
E_G=-{1\over2}\sum_{1\le i\ne j\le L}{1\over (r_i-r_j)^2} =-{1\over4}L(L-1)\ .
\end{equation}
While this is the ground state of the system in the complete
Hilbert space, we now wish to
investigate the model in a given subspace of fixed occupation numbers
$(\{Q_\alpha\},\{M_\sigma\})$.  Based on our previous experience with the
supersymmetric $t$-$J$ model and the extended Hubbard model, we anticipate
the ground state wavefunction in this sector to take the following form:
\begin{equation}
\phi_0 
(x\sigma,y\alpha)=\prod_{1\le i <j \le M} (x_i-x_j)^{\delta_{\sigma_i\sigma_j}}
e^{{i\pi\over 2} {\rm sgn}(\sigma_i-\sigma_j)}\ .
\label{eq:ground1} 
\end{equation}
This is essentially the minimal possible wavefunction that satisfies the
appropriate anti-symmetries under fermion exchange.  Using the techniques of
\cite{wang1,wang2}, we may prove that this Jastrow wavefunction is an
eigenstate of the Hamiltonian (\ref{eq:phihamil}) with eigenenergy
\begin{equation}
E_0=-{1\over 4} L(L-1) + {1\over 2} \sum_{\sigma=1}^n M_\sigma (M_\sigma -1)\ ,
\label{eq:eigenener}
\end{equation}
independent of bosonic species, as anticipated in the previous section.
Minimizing this energy subject to the constraint $\sum_\sigma M_\sigma=M$
gives
\begin{equation}
E(Q,M)=-{1\over4}L(L-1)+{1\over2n}[M(M-n)+r(n-r)]\ ,
\end{equation}
where $r=M\,{\rm mod}\,n$.  The ground state wavefunction has the fermions
distributed as evenly as possible among the $n$ distinct flavors.  For
$M\ge n$, this is the highest component of the supermultiplet.

We now turn to the question of excitations above the ground state.  As in
the supersymmetric $t$-$J$ model and the supersymmetric extended Hubbard
model, one can show that there are several ways to create excitations from
this ground state. The first way is to excite the $M$ fermions, the second
way is to excite the $Q$ bosons.  In fact, $a_k^\dagger$ and
$a_k^{\vphantom{\dagger}}$ are
raising and lowering operators of the Hamiltonian $H$\cite{wang1,wang2}. 
To construct excited state wavefunctions, we first use the convention
implicit in (\ref{eq:amplitude}) that the fermions are ordered before the
bosons.  Thus the positions of the fermions are $(q_1, q_2,\cdots,q_M)$
and the positions of the bosons are $(q_{M+1}, \cdots, q_L)$. 
One way to excite those $f$ fermions is to construct the following
wavefunctions:
\begin{equation}
|K_1,K_2,\cdots, K_M\rangle= \sum_{P_1,P_2,\cdots,P_M} \prod_{i=1}^M
(a_i^\dagger \sigma_i)^{K_{P_i}} |\phi_0\rangle\ ,
\end{equation} 
where the summation $P$ is over all possible permutations, $K_i$
are nonzero integers, and $|\phi_0\rangle$ is the state vector corresponding 
to the amplitude of Eqn.~(\ref{eq:ground1}). 
These states, if not vanishing, will be eigenstates
of the Hamiltonian, with eigenenergies given by
\begin{equation}
E(K_1, \cdots, K_M)=E_0 +\sum_{i=1}^M K_i\ .
\end{equation} 
To excite the bosons, one way is to construct the state vectors: 
\begin{equation}
|K_1,K_2,\cdots,K_Q\rangle=\sum_P \prod_{i=1}^{Q} 
(a_{i+M}^\dagger\alpha_i)^{K_{P_i}}
|\phi_0\rangle\ ,
\end{equation}
with eigenenergies given by
\begin{equation}
E=E_0+\sum_{i=1}^Q K_i\ ,
\end{equation}
if the state vectors constructed this way do not vanish.
There are also some other ways to excite the fermions and bosons, such as 
\begin{equation}
|K_1,K_2,\cdots,K_Q\rangle=\sum_P \prod_{i=1}^Q 
(a_{i+M}^\dagger)^{K_{P_i}}
|\phi_0\rangle\ .
\end{equation}

Now, let us return to the question of why $\phi_0$ is anticipated to be
the ground state of the system.  Heuristically, we note that, by
construction, $\phi_0$ is as symmetric as possible while still compatible
with fermion statistics.  Thus it is as close as possible to the maximally
symmetric state $\phi_G$.  More rigorously, we can prove explicitly that
\begin{equation}
(\sum_{i=1}^Q a_{i+M}\alpha_i) |\phi_0\rangle=0, \qquad
(\sum_{i=1}^Q a_{i+M})|\phi_0\rangle=0\ .
\end{equation} 
Furthermore, one can show that 
\begin{equation}
(\sum_{i=1}^M a_i \sigma_i) |\phi_0\rangle=0, \qquad
(\sum_{i=1}^M a_i) |\phi_0\rangle=0 \ ,
\end{equation}
for the fermions' degrees of freedom. 
These relations yield the impossibility of constructing
non-vanishing states of energy $E_0-1$ by using the 
lowering operators $a_k$.
We therefore can regard these identities as partial confirmation
that $\phi_0$ is the ground state.  A more complete proof requires further
work.

In the subspace of fixed $(\{Q_\alpha\},\{M_\sigma\})$,
the full energy spectrum of the $SU(m|n)$ model on this non-uniform
lattice is expected to consist of equally-spaced energy levels:
\begin{equation}
E_S=E_0+S\ ,
\end{equation}
where $S=0,1,2,\cdots,S_{max}$. There is an upper bound on the values
of $S$ due to the finite size of the Hilbert space.  Using the
boson--fermion interchange symmetry, $H\rightarrow -H$ for
$SU(m|n) \rightarrow SU(n|m)$, and Eqn.~(\ref{eq:eigenener}),
we find that $E_{S_{max}}= {1\over4}L(L-1)
-{1\over 2} \sum_\alpha Q_\alpha (Q_\alpha -1)$. 
Presently
we are unable to characterize the pattern of the energy spectrum by some
systematic rule, nor are we able to explain the degeneracies of the energy
levels by the underlying symmetries of the system (presumably Yangian
symmetry).

\section{{$SU(m|n)$} electronic model on a uniform lattice}

We now turn to the $SU(m|n)$ supersymmetric electronic models on a uniform
lattice in one dimension.  This uniform lattice is characterized by the
trigonometric interaction
\begin{equation}
d(i,j)={L\over\pi}\sin\left({\pi|i-j|\over L}\right)\ .
\end{equation}
The general $SU(m|n)$ model reduces to the
Haldane-Shastry spin chain \cite{haldane,shastry} for the case of
$SU(0|2)$, the supersymmetric $t$-$J$ model introduced by 
Kuramoto and Yokoyama \cite{kura,kawa,wang3} for $SU(1|2)$, and
the extended Hubbard model studied by us previously \cite{liu} for $SU(2|2)$.

For the Hamiltonian (\ref{eq:phihamil}) on a uniform lattice, we may use
the results of Fowler and Minahan for the Haldane-Shastry spin chain
\cite{fowler}, except that one has to use the permutation properties of the
amplitude $\phi$ as defined by Eqn.~(\ref{eq:amplitude}).  The result is a
set of conserved quantities $I_A$,
\begin{equation}
[H, I_A]=0,\qquad[I_A,I_B]=0,\qquad A,B=0,1,2,\cdots,\infty\ ,
\end{equation}
where
\begin{eqnarray}
&&I_A=\sum_{i=1}^L \pi_i^A\nonumber\\
&&\pi_j=\sum_{k(\ne j)=1}^L {Z_k\over(Z_k-Z_j)}M_{jk},
\qquad Z_k=\exp({2\pi i q_k\over L})\ ,
\end{eqnarray}
which yield the integrability of the $SU(m|n)$ electronic model.

In the full Hilbert space, the ground state (as the lowest component of a
supermultiplet) is obtained when $M=0$, with
trivial ground state wavefunction (\ref{eq:fullgs}).  The corresponding
ground state energy on the uniform lattice is
\begin{equation}
E_G/(\pi/L)^2=-{1\over2}\sum_{1\le i\ne j\le L}{1\over\sin^2(\pi(i-j)/L)}
 =-{1\over6}L(L^2-1)\ ,
\end{equation}
with degeneracy of the lowest component given by the $([L^1],[\cdot])$
representation of $SU(m)\times SU(n)$.

Once again, we turn to the examination of the ground state in a given
subsector specified by $(\{Q_\alpha\},\{M_\sigma\})$.  As before, when
fermions are present, we expect the ground state wavefunction to have
the minimum antisymmetry possible consistent with fermi statistics.
This time, however, due to the uniform lattice, we also require the
wavefunctions to have the appropriate periodicity under $q_i\to q_i+L$.
The results for the $SU(1|n)$ case\cite{hahaldane} motivate us to write
the following Jastrow wavefunctions for the $SU(m|n)$ case:
\begin{equation}
\phi_0(x\sigma,y\alpha) = \prod_{i=1}^M X_i^{J_{\sigma_i}}
\prod_{1\le i <j \le M}
(X_i-X_j)^{\delta_{\sigma_i\sigma_j}} e^{{i\pi \over 2} {\rm sgn}
(\sigma_i-\sigma_j)}\ ,
\label{eq:ground2}
\end{equation}
where $X_k=\exp({2\pi i x_k\over L})$, and $M=\sum_\sigma M_\sigma$ is the 
total number of fermions.  These wavefunctions are eigenstates of the
Hamiltonian provided the quantum numbers $J_1,J_2, \cdots, J_n$ (uniform
lattice momenta for each fermion flavor) obey the constraints
$-M_\sigma\le J_\sigma \le 0$.

The eigenenergies corresponding to the above wavefunctions were derived
in \cite{hahaldane}, and take on a more concise form when expressed as a
function of the shifted momenta, $K_\sigma=J_\sigma+{1\over2}(M_\sigma-1)$.
Without loss of generality, we assume the occupation numbers,
$\{M_\sigma\}$, are ordered according to $M_1\ge M_2\ge\cdots\ge M_n$.
In this case, we find
\begin{eqnarray}
E_0&=&-{1\over6}L(L^2-1)+{1\over6}\sum_{\sigma=1}^n
    (3L+(\sigma-2)M_\sigma)(M_\sigma^2-1)
-{1\over2}\sum_{\sigma<\sigma'}M_\sigma(M_{\sigma'}^2-1)\nonumber\\
&&+2(L-M)\sum_{\sigma=1}^nK_\sigma^2
+2\sum_{\sigma<\sigma'}M_{\sigma'}(K_{\sigma'}-K_\sigma)^2\ ,
\label{eq:jastrowE}
\end{eqnarray}
provided the (either integer or half integer) $K_\sigma$'s satisfy
\begin{eqnarray}
|K_\sigma|&\le&{1\over2}(M_\sigma-1)\nonumber\\
|K_\sigma-K_{\sigma'}|&\le&{1\over2}(M_\sigma-M_{\sigma'})
    \quad{\rm for}\ \sigma'>\sigma\ .
\end{eqnarray}
The lattice momentum carried by this state (over the bosonic background) is
simply $P_0=\sum_\sigma M_\sigma K_\sigma$.

Examination of Eqn.~(\ref{eq:jastrowE}) indicates that the lowest energy
state in the above set of Jastrow wavefunctions is reached when the
$K_\sigma$'s are all as close to 0 as possible.  This is accomplished by
taking $K_\sigma=0$ for odd $M_\sigma$ or $1/2$ for even $M_\sigma$.  Note
that whenever some of the $M_\sigma$'s are even, we could equally well have
taken $K_\sigma=-1/2$, leading to a two-fold degeneracy of the ground state
arising from the reflection symmetry of the lattice (in addition to the
degeneracy arising from the bosonic species).  As in the case of the
non-uniform lattice, the ground state in a given $(Q,M)$ sector has the $M$
fermions distributed as evenly as possible.  The ground state energy is then
given by
\begin{eqnarray}
E(Q,M)/(\pi/L)^2&=&-{1\over6}L(L^2-1)+{1\over6n}(3L-M)(M^2-n^2)\nonumber\\
&&+{1\over6n}r(n-r)[3(L+M)+2(n-2r)]\nonumber\\
&&+{1\over2}(L-M)\cases{n-r,&if Int($M/n$) is even;\cr r,&otherwise.\cr}
\end{eqnarray}
As before, $r$ is given by $r=M\,{\rm mod}\,n$.  Only the first line is
important in the thermodynamic limit.  This ground state (and its
reflected pair for $P\ne0$ or $L/2$) transforms as the $([Q^1],[1^r])$ of
$SU(m)\times SU(n)$ (once again the highest component of a supermultiplet
for $M\ge n$).  We note that when
all $M_\sigma$ are odd, this particular wavefunction can also  
be obtained by taking the strong interaction limit of the corresponding
continuum quantum system\cite{kato}.  

The full energy spectrum for this long range permutation model may be
derived using the asymptotic Bethe-ansatz (ABA) \cite{sutherland}.  The
motivation for this ABA lies in the fact that the scattering is
essentially two-body in nature, even in the presence of long range
interactions, as indicated by the Jastrow form of the wavefunction.  The
ABA has been used successfully in the $SU(1|2)$ supersymmetric $t$-$J$
model \cite{kawaaba} as well as its $SU(1|n)$ generalization \cite{kawa},
and gives {\it exact} results, even in the non-asymptotic regime,
as proven to be true for the $SU(0|2)$ 
Haldane-Shastry spin chain\cite{haldane2} and 
for the $SU(1|2)$ $t$-$J$ model\cite{wang3} --- a fact
that is related to the integrability of the system.

For the $SU(m|n)$ generalization of the ABA, we need to treat multiple
flavors of both bosons and fermions on the lattice.  In the case of
nearest-neighbor interactions, Lai first discussed such lattice permutation
models with mixtures of bosons and fermions in 1974\cite{lai}.  This was
later generalized by various authors\cite{sutherland2,sto,korepintj,korepin3}.
For the multi-species case, the Bethe-ansatz takes on a nested form, where
one species is removed at each step.  For the long range $SU(m|n)$ model,
this is accomplished by starting with one of the $m+n$ (either bosonic or
fermionic) species as the background, and then peeling off the remaining
$m+n-1$ species one at a time, resulting in $m+n-1$ sets of Bethe-ansatz
equations.  There are a total of $(m+n)!$ ways to perform this nesting.
However this number may be reduced by the obvious $n!$ fermion permutations
and $m!$ boson permutations, yielding $(m+n)!/(m!n!)$ independent ways of
writing the nested Bethe-ansatz.  A particular choice of
nesting may be denoted by a string of $m+n$ $B$'s or $F$'s, indicating the
order in which the bosonic and fermionic species are removed, working from
right to left (the rightmost character denotes the choice of background).
Of course all such choices of the nesting should yield equivalent results.
The ABA obtained previously for the $SU(1|n)$ supersymmetric $t$-$J$ model
\cite{kawa} corresponds to the $BF^n$ choice of nesting.

For a general $\nu$ ($=m+n$) component system, we encode the nesting of the
ABA according to the statistics $s_i$ of species $i$  where $s_i=0$ for
bosons and 1 for fermions.  Since we take the first species as the
background, the $B^mF^n$ nesting corresponds to
$\{s\}=(1,\ldots,1,0,\ldots,0)$.  We denote the occupancies of the species
by $M_i$ (here $i$ runs from 1 to $\nu$) and define the quantities
\begin{equation}
N_i=\sum_{j>i}^\nu M_j\ ,
\end{equation}
where $N_0=L$ and $N_\nu=0$.  For the nested ABA, we introduce $\nu-1$
sets of pseudomomenta, $\{p^{(1)}\},\{p^{(2)}\},\ldots,\{p^{(\nu-1)}\}$
where each set $\{p^{(a)}\}$ consists of the $N_a$ quantities
\begin{equation}
p_i^{(a)}:\quad i=1,2,\ldots,N_a\ ,
\end{equation}
all of which are within the range $[-\pi,\pi]$.  The energy and the total
lattice momentum of the system depend only on the $p_i^{(1)}$'s, and
are given by
\begin{eqnarray}
&&E=(-1)^{s_1+1}\left[{\pi^2\over6}L(1-{1\over L^2})+
{1\over 2} \sum_{i=1}^{N_1} ((p_i^{(1)})^2-\pi^2)\right]\nonumber\\
&&P=\left[s_1(L-1)\pi + \sum_{i=1}^{N_1} (p_i^{(1)}-\pi)\right]
\ {\rm mod}\ 2\pi\ .
\label{eq:abaenergy}
\end{eqnarray}

The pseudomomenta, $\{p^{(a)}\}$ for $a=1,\ldots,\nu-1$, are obtained by
solving the $\nu-1$ set of equations of the nested ABA, expressed concisely
in the following form:
\begin{equation}
\sum_{k=1}^{N_{a-1}}\theta(p_i^{(a)}-p_k^{(a-1)})
=2\pi I_i^{(a)}+\delta_{s_a,s_{a+1}}\sum_{j=1}^{N_a}\theta(p_i^{(a)}-p_j^{(a)})
+(-1)^{\delta_{s_a,s_{a+1}}}\sum_{l=1}^{N_{a+1}}\theta(p_i^{(a)}-p_l^{(a+1)})
\ ,
\label{eq:aba}
\end{equation}
where $\theta(x)=\pi {\rm sgn}(x)$ is a step function.  For $a=1$ the left
hand
side of (\ref{eq:aba}) is replaced by $p_i^{(1)}L$, and for $a=\nu-1$ the
last term on the right hand side is dropped.  A given state in the ABA
spectrum is thus specified by the set of quantum numbers $\{I^{(a)}\}$.  At
a given level $a$ of the nesting, there are $N_a$ non-overlapping
$I_i^{(a)}$'s (all integers or half-integers as appropriate) which are
required to lie in the range
\begin{equation}
|I_i^{(a)}|\le{1\over2}(N_{a-1}-N_a+N_{a+1}-1)\ ,
\label{eq:cond1}
\end{equation}
for $s_a=s_{a+1}$ or
\begin{equation}
|I_i^{(a)}|\le{1\over2}(N_{a-1}-N_{a+1}-2)\ ,
\label{eq:cond2}
\end{equation}
for $s_a\ne s_{a+1}$.  These conditions lead to the restrictions on the
occupation numbers, either $M_a\ge M_{a+1}$ or $M_a\ge N_{a+1}+1$
respectively.  An exception happens when $N_a=0$, in which case the nested
ABA terminates early, giving no further restriction on $M_a$.

We remark that in general the ABA still provides exact eigenenergies
when conditions (\ref{eq:cond1}) and (\ref{eq:cond2}) are relaxed by
one unit on either end.  However we have found the above restrictions to be
necessary in order to ensure that the ABA generates only highest weight
states of the $SU(m|n)$ superalgebra.  It is important to realize that the
choice of nesting plays a crucial role in defining what is meant by the
highest weight representation.  In particular, for $SU(m|n)$, the Cartan
subalgebra lies in the bosonic subgroup $SU(m)\times SU(n)\times U(1)$ and
has rank $m+n-1$.  Thus highest weight representations of $SU(m|n)$ are
annihilated by a set of $m+n-1$ simple positive roots.  It is this set of
simple roots which is determined by the nesting of the ABA, with identical
statistics in the nesting ($s_a=s_{a+1}$) corresponding to bosonic roots and
$s_a\ne s_{a+1}$ corresponding to fermionic roots.  For this reason the most
straightforward nesting is given by $B^mF^n$ (or $F^nB^m$), in which case
there are $m+n-2$ bosonic roots corresponding to the simple positive roots
of $SU(m)\times SU(n)$ and a single fermionic root taking a state from the
$(Q,M)$ sector to $(Q-1,M+1)$.  This gives the conventional definition of
the highest weight state of $SU(m|n)$ residing in the highest component of
the supermultiplet.

In order to be more specific, we now pick the natural $B^mF^n$ nesting and
the original notation of $M_\sigma$ and $Q_\alpha$ for fermionic and bosonic
occupation numbers respectively.  The nested ABA may be written in terms of
the $n-1$ sets of ``fermionic'' pseudomomenta
\begin{equation}
p_i^{(a)}:\qquad i=1,2,\ldots, N_a
\quad {\rm where}\quad N_a=Q+\sum_{\sigma=a+1}^nM_\sigma\ ,
\end{equation}
and $m$ sets of ``bosonic'' pseudomomenta
\begin{equation}
q_\mu^{(b)}:\qquad \mu=1,2,\ldots,N'_b
\quad {\rm where}\quad N'_b=\sum_{\alpha=b}^m Q_\alpha\ .
\end{equation}
The ABA may now be written explicitly as
\begin{eqnarray}
&&p_i^{(1)} L=2\pi I_i^{(1)} +\sum_{i'(\ne i)} \theta(p_i^{(1)}-p_{i'}^{(1)})
-\sum_j \theta(p_i^{(1)}-p_j^{(2)})\ ,\nonumber\\ 
&&\sum_i \theta(p_j^{(2)} - p_i^{(1)})
=2\pi I_j^{(2)} + \sum_{j'(\ne j)} \theta (p_j^{(2)} -p_{j'}^{(2)})
- \sum_k \theta (p_j^{(2)} -p_k^{(3)})\ ,\nonumber\\
&&\ \,\vdots\nonumber\\
&&\sum_j \theta(p_k^{(n-1)} - p_j^{(n-2)})
=2\pi I_k^{(n-1)} + \sum_{k'(\ne k)} \theta (p_k^{(n-1)} -p_{k'}^{(n-1)})
- \sum_\mu \theta (p_k^{(n-1)} -q_\mu^{(1)})\ ,\nonumber\\
%
%
&&\sum_k \theta(q_\mu^{(1)}-p_k^{(n-1)})=2\pi J_\mu^{(1)} + 
\sum_\nu \theta (q_\mu^{(1)}-q_\nu^{(2)})\ ,\nonumber\\
&&\sum_\mu \theta(q_\nu^{(2)}-q_{\mu}^{(1)})
= 2\pi J_\nu^{(2)} + \sum_{\nu'(\ne\nu)} \theta (q_\nu^{(2)}-q_{\nu'}^{(2)})
-\sum_\gamma \theta(q_\nu^{(2)}-q_\gamma^{(3)})\ ,\nonumber\\
&&\ \,\vdots\nonumber\\
&&\sum_\nu \theta(q_\gamma^{(m-1)}-q_{\nu}^{(m-2)})
= 2\pi J_\gamma^{(m-1)} + \sum_{\gamma'(\ne\gamma)}
\theta (q_\gamma^{(m-1)}-q_{\gamma'}^{(m-1)})
-\sum_\eta \theta(q_\gamma^{(m-1)}-q_\eta^{(m)})\ ,\nonumber\\
&&\sum_\gamma\theta (q_\eta^{(m)} - q_\gamma^{(m-1)})
=2\pi J_\eta^{(m)} + \sum_{\eta'(\ne\eta)}
\theta (q_\eta^{(m)} - q_{\eta'}^{(m)})\ ,
\end{eqnarray}
with quantum numbers $\{I^{(a)}\}$ and $\{J^{(b)}\}$.  
It is anticipated that the ABA
span the full set of energy levels when both the highest weight
property of the ABA states and the supermultiplet structure connecting
states in different $(Q,M)$ sectors are taken into account.  We have
verified that this is the case numerically for the $SU(2|2)$ model
on small lattices.  The numerical results also indicate that, while
the ABA accounts for the full energy spectrum, it does not give all
the proper degeneracies of the states.  The missing states presumably
arise from localized bound states which are not described by the
{\it asymptotic} Bethe-ansatz.  Thus, in order to properly count the
degeneracies and derive the full $SU(m|n)$ supermultiplet structure of the
spectrum, one must generalize the ``squeezed string'' picture of the $SU(n)$
Haldane-Shastry spin chain\cite{hahaldane2}.  
 
\section{Conclusion}

In this work, we have introduced the $SU(m|n)$ electronic model of long
range interaction, which is integrable on both uniform and non-uniform one
dimensional lattices.  The ground state, excitation spectrum, and
integrability properties of this model have been considered in detail.
Since the $SU(m|n)$ model includes many previously well-known models as
special cases, we have shown how a uniform treatment of all such models may
be accommodated.

For the integrable $SU(m|n)$ model on a non-uniform lattice, perhaps
the most interesting open problem is to explain the high degeneracy
of the equally-spaced energy levels using the underlying symmetries of
the system.  Presently it is still unclear how to construct the
thermodynamics of the long range $SU(m|n)$ permutation model on the
Hermite lattice.  

For the uniform lattice, since the energies of the ABA states are given
by the essentially non-interacting formula, (\ref{eq:abaenergy}), we note
the interesting result that for a given lattice of length $L$, the set
of allowed energy levels are completely determined, and is independent of
the number of fermionic and bosonic species, $n$ and $m$ \cite{haldanetalk}.
The differences between the $SU(m|n)$ models on a uniform lattice hence
lie only in the degeneracies and supermultiplet structures of the models,
with larger $n$ and $m$ leading to higher degeneracies.

Since the full $SU(m|n)$ symmetry is broken in most physical problems, 
we generally wish to work in a given subspace of fixed $(Q,M)$,
corresponding to the breaking of supersymmetry.  It is
clear that, in this case, a proper understanding of both the supermultiplet
structure (including degeneracies) and $SU(m|n)$ representation theory is
required.  We note that the latter is quite intricate as atypical
representations are clearly involved.

Working with fixed $(Q,M)$, one only sees the bosonic $SU(m)\times SU(n)$
subgroup of the full $SU(m|n)$ symmetry group.  For both the non-uniform and
uniform lattice, we have explicitly found the ground state wavefunctions
and their corresponding eigenenergies.  Because of the symmetry properties
of the wavefunction, the ground state wavefunction is independent of the
bosons.  Hence the ground state in any $Q>0$ sector is degenerate whenever
$m>1$, and transforms in the $([Q^1],[1^r])$ representation where
$r=M\,{\rm mod}\,n$.

The ground state structure and ABA results for the excited spectrum of
states on the uniform chain presented above opens up the possibility of
studying the properties of elementary excitations of 
all the general $SU(m|n)$
models in a consistent manner.  While we have in particular focused on
lattice models with inverse square exchange, our results carry over to
many other graded permutation models as well.

\bigskip

This work was supported in part
by the U.~S.~Department of Energy under grant no.~DOE-91ER40651-TASKB,
and by the Swiss National Science Foundation.

\end{document}